\documentstyle[prl,aps]{revtex}

\newcommand{\ai}{\'{\i}}
\begin{document}

\draft
\title{Quantum effects in the scattering by the magnetic 
       field of a solenoid.} 
\author{Gabriela Murgu\ai{}a \footnote{E-mail: gabriela@ft.fisica.unam.mx} 
        and Mat\ai{}as Moreno \footnote{E-mail: matias@fisica.unam.mx}}
\address{Instituto de F\ai{}sica, Universidad Nacional Aut\'onoma de M\'exico.\\
         Apartado Postal 20364, 01000, M\'exico, D.F. M\'exico.}  
\date{\today}

\maketitle 

%%% Abstract %%%
\begin{abstract}
  
  We present a relativistic quantum calculation at first order in
  perturbation theory of the differential cross section for a Dirac
  particle scattered by the magnetic field of a solenoid. The
  resulting cross section is symmetric in the scattering angle as
  those obtained by Aharonov and Bohm~(AB) in the string limit and by
  Landau and Lifshitz~(LL) for the non relativistic case.  We show
  that taking $pr_0\sin{(\theta/2)}/\hbar\ll 1$ in our
  expression of the differential cross section it reduces to that one
  reported by AB, and if additionally we assume $\theta \ll 1$ our
  result becomes the one obtained by LL. However, these limits are
  explicitly singular in $\hbar$ as opposed to our initial result. We
  analyse the singular behavior in $\hbar$ and show that the
  perturbative Planck limit ($\hbar \rightarrow 0$) is consistent,
  contrarily to those of the AB and LL expressions.
\end{abstract}

We know that in the classical scattering of charged particles by
magnetic fields the particles describe circular trajectories
with fixed radii, so they have a preferential movement direction. In
this Letter, the relativistic quantum version of the problem is
studied in the lowest order of perturbation theory and a symmetric
behavior in the scattering angle is found for a solenoidal magnetic
field. We study some interesting limit cases of the differential cross
section and compare them with previous non relativistic results
reported by Aharonov and Bohm (AB)~\cite{AB} and Landau and Lifshitz
(LL)~\cite{LL}.  

As it is known, the Aharonov-Bohm effect~\cite{AB} is considered one
of the most important confirmed~\cite{Chambers-Tonomura} predictions
of quantum mechanics because it shows that the vector potential has a
physical significance and can be viewed more than a mathemathical
convenience. The interest in this effect has been increased
recently~\cite{Varios}. Both because of basic reasons that have
changed the understanding of gauge fileds and forces in nature and
also because it has a lot of connections with new physics, like the
quantum Hall effect~\cite{QHE}, mesoscopic physics~\cite{MesoscopicP}
and physics of anyons~\cite{Anyons}.

We want to point out that althought the present calculation seems to
be elementary, indeed it is a calculation that involves certain
technical problems, as we will show. In this work, we will follow the
Bjorken and Drell convention~\cite{BD}.

Let us consider the scattering of a Dirac particle by the magnetic
field of a solenoid with a constant magnetic flux. Note that this is a
problem in which free particle asymptotic states can be used.  Also
notice that the global phase $\exp{(ie\Phi\theta/hc)}$ in the free
particle wave function (because of the presence of a pure gauge field
in the exterior of the solenoid) does not contribute to the $S$ matrix.

Consider a long solenoid of lenght $L$ and radius $r_0 \ll L$ centered
along the ${\bf \hat{\i}_3}$ axis. Inside of the solenoid, where $r<r_0$,
the magnetic field is uniform, ${\bf B}=B_0{\bf \hat{\i}_3}$, with
$B_0$ being a constant, while outside of the solenoid, where $r>r_0$,
the magnetic field is null. Replacing the vector potential
that describes this magnetic field in the lowest order in $\alpha$ of
the $S$ matrix for Dirac particle solutions and multiplying by the
phase space factors, we obtain the differential cross section per unit
length of the solenoid:
\begin{equation} 
\frac{d\sigma}{dx_3 d\theta} = 
     \frac{1}{f} \frac{\hbar}{c^2} \left({\frac{e\Phi}{r_0}}\right)^2 
     \frac{{\left| 
              J_1(2\frac{p}{\hbar}r_0\left|\sin{\frac{\theta}{2}}\right|)                  \right|}^2}
          {8\pi p^3 \sin^4{\frac{\theta}{2}}},
\label{dsigma} 
\end{equation} 
which has the same form whether or not the final polarization of the
beam is actually measured ($f=1$ or $f=2$), so this result does not
depend on the final polarization. In Eq.~(\ref{dsigma}), $J_1$ are the
first order Bessel functions~\cite{Arfken}, $\theta$ is the scattering
angle, $p$ stands for the momentum of the incident particle and 
$\Phi = \pi r_{0}^{2} B_0$ is the total magnetic flux.

As it can be observed, the differential cross section is symmetric in
$\theta$. This is reminiscent of the Stern-Gerlach result, in which an
unpolarized beam interacting with an inhomogeneous magnetic field is
equally split into two parts, each one with opposite spin. But, as we
have mentioned before, Eq.~(\ref{dsigma}) does not depend on the final
polarization of the particles.  Thus, this symmetric behavior of
$\theta$ should be a consequence of the perturbation theory, althought
this symmetry is also present in non perturbative results~\cite{AB,LL}.

We want to study the limit case of small scattering angles.  So, if we
assume $pr_0\sin{(\theta/2)}/\hbar\ll 1$, then the cross
section of Eq.~(\ref{dsigma}) reduces to
\begin{equation}
\left.{\frac{d\sigma}{dx_3 d\theta}}
      \right|_{\frac{p}{\hbar}r_0\sin{\frac{\theta}{2}} \ll 1} = 
        \frac{1}{f} 
        \frac{e^2\Phi^2}{8\pi c^2 \hbar p  \sin^2{\frac{\theta}{2}}},
\label{secc.mm.r}
\end{equation}
which agrees with the result reported by Aharonov and Bohm~\cite{AB} when
$e\Phi/2\hbar c \ll 1$. If we additionally impose the
condition $\theta \ll 1$ we obtain
\begin{equation}  
\left.{\frac{d\sigma}{dx_3 d\theta}}  
    \right|_{\frac{p}{\hbar}r_0\sin{\frac{\theta}{2}} \ll 1, \theta \ll 1}   
  = \frac{1}{f} \frac{e^2\Phi^2}{2\pi c^2 \hbar p \theta^2},
\label{secc.mm.h.c}  
\end{equation}  
which is precisely the result reported by Landau and Lifshitz~\cite{LL}.
  
Note that the cross section is singular when $\hbar \rightarrow 0$ for
both limiting cases resulting in a classical limit that apparently
diverges. We want to point out that it does not make sense to take the
Planck's limit ($\hbar \rightarrow 0$) in Eq.~(\ref{secc.mm.r}) or in
Eq.~(\ref{secc.mm.h.c}), because both expressions were obtained
assuming the condition $pr_0\sin{(\theta/2)}/\hbar \ll 1$. Hence, we
have to take the classical limit using the expression for the
differential cross section given in Eq.~(\ref{dsigma}).  Defining 
$x = 2pr_0{\left|\sin{(\theta/2)}\right|}/\hbar = r_0q$, we observe 
that the limit $\hbar \rightarrow 0$ implies $x \rightarrow \infty$ or
$pr_0 \rightarrow \infty$~\cite{Zkarzhinsky-97} for fixed $\theta$.
Using the asymptotic behavior of the Bessel function~\cite{Arfken}

$$
\lim_{x \rightarrow \infty}{J_1(x)} = 
      - \sqrt{\frac{2}{\pi x}}\cos{\left(x - \frac{3}{4}\pi\right)},
           \hspace{.5cm} x \gg \frac{3}{8};
$$
the resulting Planck classical limit of Eq.(\ref{dsigma}) is
identically zero for fixed $e, p, r_0, \Phi,$ and $\theta$:
\begin{equation}
\lim_{\hbar \rightarrow 0}\frac{d\sigma}{dx_3 d\theta}
= \lim_{\hbar \rightarrow 0}
  \frac{\hbar^2}{f} \left({\frac{e\Phi}{2\pi c}}\right)^2
     \frac{\cos^2{\left({2\frac{p}{\hbar}r_0 
                        \left|\sin{\frac{\theta}{2}}\right| - 
                        \frac{3}{4}\pi}
                  \right)}
          }
          {2 r_{0}^{3} p^4 \left|{\sin^5{\frac{\theta}{2}}}\right|}
= 0 
\label{climit-dsigma}
\end{equation} 
and does not show any singularity in $\hbar$ as compared to those of
small scattering angles, Eq.~(\ref{secc.mm.h.c}), or small radii of
the solenoid, Eq.~(\ref{secc.mm.r}). 
Notice that the perturbative result gives a consistent finite
classical limit and reduces to the eikonal and the zero radius
limits and it  shows that taking the classical limit of such
results is misleading. So, it is worthwhile to notice
that the order in which the limits are taken is crucial.

The apparent difference in the classical limits comes from the fact
that in taking the limit $\hbar \rightarrow 0$ of the perturbative
result, Eq.~(\ref{dsigma}), the Bessel function decreases as $J_1(x)
\sim 1/\sqrt{x}$ and it does generate an $\hbar$ contribution to the
cross section. On the other hand, if one begins by taking the small
angle or the small radius limit, the Bessel function approximates to
$J_1(x) \sim x$, and this behaves like $1/\hbar$. The overall
difference between these two procedures is an $\hbar^3$ factor. It is
important to notice that loop corrections to the perturbative
expansion do not modify the $\hbar$ behavior of the physical
amplitude, as can be proved with the use of the loop expansion.

One can obtain a non divergent expression for the Landau and Lifshitz
and the Aharonov and Bohm results when $\hbar \rightarrow 0$ if one
quantizes $\Phi$, the magnetic flux. Imposing the magnetic flux
quantization condition, $\Phi = n\Phi_0$, where $\Phi_0 = hc/e = 4.318
\times 10^{-7}$~gauss~cm$^2$, the cross section of Eq.~(\ref{dsigma})
takes the form
$$
\frac{d\sigma}{dx_3 d\theta} = 
    n^2 \hbar^3 \frac{\pi}{f}  
    \frac{{\left| 
            J_1(2\frac{p}{\hbar}r_0\left|\sin{\frac{\theta}{2}}\right|)
           \right|}^2}
         {2 r_0^2 p^3 \sin^4{\frac{\theta}{2}}},
$$
which apart of being independent of the charge of the particles, it is
a cross section of a purely quantum effect.  Cast in this way there is
no singular behavior in $\hbar$, in contradistinction to the form that
Landau and Lifshitz report. In particular, for the case of small
scattering angles, it takes the form
$$
\left.{\frac{d\sigma}{dx_3 d\theta}}\right|_{\theta \ll 1} = 
          \frac{8 \pi^2 \hbar n^2}{f p \theta^2},
$$
which also has a null classical limit. 

Finally, we want to point out that althought our reslut is consistent
in the sense that the Aharonov and Bohm and the Landau and Lifshitz
results are recovered, there is no direct classical correspondence via
the Planck's limit (see Eq.~(\ref{climit-dsigma})), because in
particular the cross section is symmetric in $\theta$. This problem is
shared also by the AB and LL solutions and is possibly solved by higher
order corrections in the {\it external} magnetic field.

%%% Acknowledgments %%%
\section*{Acknowledgments}

We want to thank the helpful comments of A.~Rosado.  This work was
partially supported by CONACyT~(3097P-E), DGAPA-UNAM~(IN118600) and
DGEP-UNAM.

%%% References %%%

\end{document}